\documentclass[doublecol]{epl2}
\usepackage{amsmath}
\usepackage{amsfonts}
\usepackage{graphicx}
\usepackage{amssymb}
\usepackage{float}

\title{A Green's function  approach to the Casimir effect  on topological insulators with planar symmetry}
\author{A. Mart\'{i}n-Ruiz \inst{1}, M. Cambiaso \inst{2} \and L. F. Urrutia \inst{1}} \institute{
\inst{1} Instituto de Ciencias Nucleares, Universidad Nacional Aut\'{o}noma de M\'{e}xico, 04510 M\'{e}xico, Distrito Federal, M\'{e}xico \\ \inst{2} Universidad Andres Bello, Departamento de Ciencias Fisicas, Facultad de
Ciencias Exactas, Av. Republica 220, Santiago, Chile}
\abstract{{We investigate the
Casimir stress on a topological
insulator (TI) 
between two 
metallic plates.  The TI is assumed to be  
joined to one of the plates and 
its surface in front of the other  is covered by a thin magnetic layer,
which turns the TI into a
full insulator. We also analyze the limit
where one of the plates is sent to
infinity yielding 
the Casimir stress between a
conducting plate and a TI. 
To this end we employ a local
approach in terms of the stress-energy tensor of the system,
its vacuum expectation value being subsequently evaluated
in terms of the
appropriate Green's function. 
Finally, the construction of
the renormalised vacuum stress-energy tensor
in the region between the plates  yields the 
Casimir stress.
Numerical result are also presented.}}
\begin{document}
\maketitle
\section{I. Introduction} The Casimir effect
(CE)
\cite{Casimir} is one of the most
remarkable consequences of the nonzero
vacuum energy predicted by quantum
field theory which has been confirmed by
experiments \cite{Bressi}. In its most
basic form, the CE results in the
attraction between  two perfectly
reflecting planar surfaces due to a restriction
of the allowed modes in the vacuum
between them. This attraction manifests itself when the
surfaces are separated by
a few micrometers. 
In general, the CE can be defined as the stress (force
per unit  area) on bounding surfaces
when a quantum field is confined in a
finite volume of space. The boundaries
can be  material media, interfaces between
two phases of the vacuum, or topologies
of space. For a review  see, for example, 
Refs. \cite{Milton,Bordag}. The experimental accessibility to 
micrometer-size physics has motivated
the theoretical study of the CE in
different scenarios, including the standard
model\cite{Frank} and the
gravitational sector\cite{Quach}.

The recent discovery of 3D  topological insulators (TIs)
\cite{Liang} provides an additional arena where the CE can be studied.
TIs are an emerging  class of time-reversal symmetric materials
which have attracted
much attention due to the unique
properties of their surface states.
Experimental devices with TIs are now
feasible, however the induced  topological magnetoelectric effect (TME) has not yet
been observed. In this
regard, the authors in Ref.
\cite{science,rosenberg} proposed an experimental
setup using  TIs to
measure the Witten effect. Similarly,  
it has been proposed that the
half-quantized
Hall conductances on the surfaces of two
TIs can be measured \cite{Maciejko}.
The Casimir force
between
TIs was computed in Ref. \cite{Grushin}
and the authors  proposed to measure it
using T1BiSe$_{2}$, however
the required experimental precision 
has not been achieved yet. 
This proposal also included the most notable feature that,  due to the TME, the
strength and sign of the Casimir stress
between two planar topological insulators
can
be tuned. We observe that the
calculation in Ref. \cite{Grushin} was
done by using the {scattering approach to the Casimir effect}, \textit{i.e.,} using the Fresnel coefficients for
reflection matrices at the
interfaces of the TIs. Additional  TME include: induced  mirror magnetic monopoles due to 
electric
charges close to the surface of a TI
(and vice versa) \cite{science} and a
non-trivial Faraday rotation of the
polarisations of
electromagnetic waves propagating
through a TIs surface \cite{Hehl}.

The low-energy effective
field theory (EFT) which describes the
TME, independently of microscopic
details, consists of the usual
electromagnetic Lagrangian
density supplemented by a term
proportional to
$ \theta \textbf{E} \cdot \textbf{B} $,
where $\theta$ is
the topological magnetoelectric
polarisability (TMEP) \cite{Qi}.
Time-reversal (TR) symmetry indicates
that this EFT describes the bulk
of a 3D TI when $\theta = 0$ (trivial TI)
and $\theta = \pi$ (non-trivial TI).
When the surface of the TI is included,
this theory is a fair description of both the
bulk and the surface only when a
TR breaking
perturbation is induced on the surface to
gap the surface states, thereby
converting it into a full insulator. In this
situation, which we consider
here, $\theta$ can be
shown to be quantised in odd integer
values of $\pi$:
$\theta = (2n+1) \pi$, where $n \in \mathbb{Z}$ is
determined by the nature
of the TR breaking
perturbation, which could be controlled 
experimentally by covering the TI with a
thin magnetic layer \cite{Grushin}.

In this letter we 
focus on calculating the effects of
the $\theta$-term in the Casimir stress,
restricting 
to the purely
topological contribution. The Casimir system
we consider is formed by
two perfectly reflecting planar surfaces
(labeled $P_{1}$ and
$P_{2}$) separated by a
distance $L$, with 
 a non-trivial TI placed between
them, but perfectly joined to the
plate $P_{2}$, as shown in Fig. \ref{fig1}.  The
surface $\Sigma $ of the TI, located at $z=a$, is assumed to be covered by a
thin magnetic layer which breaks TR
symmetry there.
To this end, we
follow an approach similar
to that in Ref. \cite{Brown} by performing
a local analysis
 of the forces produced by  the quantum vacuum in
$\theta$-extended electrodynamics (to
be called $\theta$-ED). In particular, 
we first construct the appropriate Green's
function (GF) for $\theta$-ED, and then
we compute the renormalised
vacuum stress-energy tensor in the region
between the plates. 
With these, we obtain 
the Casimir stress that
the plates exert on the surface $\Sigma$
of the TI. Finally we  consider  the limit
where the plate $P_{2}$ is sent to
infinity ($L \rightarrow \infty $)  to obtain the Casimir stress
between a conducting plate and a
non-trivial semi-infinite TI. 
{We take this local approach, not only
as an alternative method compared to the
scattering approach, but also as a means to illustrate yet another use of the GF 
method to unravel the electromagnetics of TIs we reported 
in Ref. \cite{MCU}, where-from we take notations and conventions.}

\section{II. Effective model of 3D TIs} \label{thetaED}

\label{FE-BC}

\begin{figure}[tbp]
\begin{center}
\includegraphics[width=2.5in]{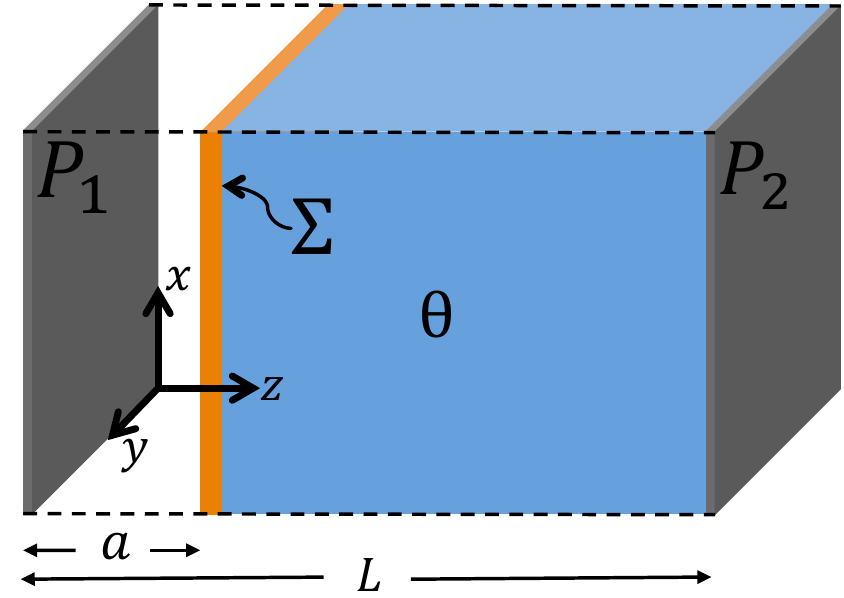}
\end{center}
\caption{{\protect\small Schematic of the Casimir effect in $\theta$-ED}.}
\label{fig1}
\end{figure}

The effective action governing the
electromagnetic response of 3D TIs,
written in a manifestly covariant way, is
\begin{equation}
\mathcal{S}=\int d^{4}x\left[ -\frac{1}{16\pi }F_{\mu \nu}F^{\mu \nu }-
\frac{\theta}{4} \frac{\alpha }{4\pi^2 }F_{\mu \nu }
\tilde{F}^{\mu \nu }-j^{\mu }A_{\mu }\right] ,  \label{action}
\end{equation}%
with $\alpha  \simeq 1 / 137$  the
fine structure constant,   
$j^{\mu }$ is   a conserved current
 coupled to the electromagnetic
potential $A _{\mu}$,   $F _{\mu
\nu} = \partial _{\mu} A _{\nu} - \partial
_{\nu} A _{\mu}$ is the 
field strength and $\tilde{F}^{\mu\nu} \equiv
\epsilon^{\mu\nu\rho\sigma}F_{\rho\sigma}/2$. The
equations of motion are
\begin{equation}
\partial _{\mu }F^{\mu \nu } + \frac{\alpha}{\pi} ( \partial _{\mu} \theta )
 \tilde{F}^{\mu \nu } = 4\pi j^{\nu } ,  \label{FieldEqs}
\end{equation}
which extend the usual Maxwell equations
to incorporate the topological
$\theta$-term.
In the
problem at hand, depicted in Fig. \ref{fig1}, we consider the standard boundary conditions (BC) for the perfectly reflecting metallic plates $P_1$
and $P_2$. 
Thus, the appropriate BC there are $n _{\mu} \tilde{F}
^{\mu \nu} \vert _{P_{1,2}}
= 0$, where $ n_{\mu} = (0,0,0,1)$.
If $\theta$ is constant
in the whole space, the
propagation of electromagnetic fields is
the same as in standard electrodynamics.
However, when the electromagnetic field
propagates through the
surface of a TI, TMEs take place. These effects are
incorporated by writing the TMEP of the TI slab
in the form
\begin{equation}
\theta (z) = \theta H (z - a) H (L - z) , \label{theta}
\end{equation}
where $H(x)$ is the Heaviside function. 
In the
Lorentz gauge $\partial _{\mu} A ^{\mu}
= 0 $,
the equation of motion for the 
potential, in the region between the plates is 
\begin{equation}
\left[ \eta ^{\mu} _{\phantom{\mu} \nu} \partial ^{2} -
\tilde{\theta} \delta \left( \Sigma \right) n _{\sigma}
\epsilon^{\sigma \mu \alpha }_{\phantom{\sigma\mu\alpha} \nu}
\partial_{\alpha } \right] A^{\nu }=4 \pi j^{\mu } . \label{FiedlEqPlaneConfig}
\end{equation}
Here $\partial ^{2} = \partial _{\mu}
\partial ^{\mu} = \partial ^{2} _{t} -
\nabla ^{2}$ and
$\tilde{\theta} = - \alpha \theta / \pi$. The boundary term
(at $z=L$), missing
in Eq. (\ref{FiedlEqPlaneConfig}),
identically vanishes in the distributional sense,  due to the BC on the
plate $P_{2}$. In this way,
Eq. (\ref{FiedlEqPlaneConfig}) implies that
the only TME present in our Casimir
system is the one produced at $\Sigma$.
Note that the field equations in
the bulk regions, vacuum $[0,a)$ and TI $(a,L]$, are the
same as in standard electrodynamics, and
that the $\theta$-term affects the
fields only at the interface $\Sigma$.
Assuming that the
time derivatives of the fields are finite in
the vicinity of $\Sigma $, together with the absence of free
sources on $\Sigma$,
Eq.
(\ref{FiedlEqPlaneConfig}) implies the
 following BC for the potential
at the interface
\begin{equation}
A^{\mu} \big| _{z= a^{-}} ^{z = a^{+}} = 0 \;\; , \;\; \left( \partial _{z}
A ^{\mu} \right) \big| _{z= a^{-}} ^{z = a^{+}} = - \tilde{\theta}
\epsilon^{3 \mu \alpha} _{\phantom{3 \mu \alpha} \nu}
\partial _{\alpha} A ^{\nu } \big| _{z=a},  \label{BC4Pot}
\end{equation}
which are derived by
integrating the field equations over a
pill-shaped region across $\Sigma$. The discontinuity
in $\partial_z A^\mu$ across $\Sigma$  produces  the
transmutation between the electric field
and the magnetic field, which characterizes
the TME of TIs.
To obtain the general solution of
Eq. (\ref{FiedlEqPlaneConfig}) for arbitrary
external
sources, we introduce the GF matrix
$G ^{\nu} _{\phantom{\nu} \sigma} \left( x , x
^{\prime} \right)$ satisfying
\begin{equation}
\left[ \eta ^{\mu} _{\phantom{\mu} \nu} \partial ^{2} -
\tilde{\theta}\delta \left( \Sigma \right)
\epsilon^{3 \mu \alpha}_{\phantom{3 \mu \alpha} \nu}
\partial _{\alpha } \right] G ^{\nu} _{\phantom{\nu} \sigma}
\left( x , x ^{\prime} \right) = 4 \pi
\eta ^{\mu} _{\phantom{\mu} \sigma}
\delta  \left( x -  x ^{\prime} \right) ,  \label{EqsGreenMatrix}
\end{equation}
together with the BCs in Eq. 
(\ref{BC4Pot}). Next we solve Eq.
(\ref{EqsGreenMatrix}) along
the same lines introduced in Ref.
\cite{MCU} for the static case. 
{A similar technique has been used in Refs. \cite{Bordag2,Milton2} to study two parallel planes represented by two $\delta$-functions.} The GF we consider has
translational invariance in the directions
parallel to $\Sigma $, that is in the
transverse $x$ and $y$ directions, while
this invariance is broken in the $z$
direction. Exploiting this symmetry we
can write
\begin{equation}
G _{\phantom{\mu} \nu }^{\mu }\left( x , x ^{\prime }\right) =
4 \pi \int \frac{d^{2}\mathbf{p}}{\left( 2\pi \right) ^{2}}
e^{i\mathbf{p} \cdot \textbf{R}} \int \frac{d \omega}{2 \pi}
e ^{-i \omega ( t - t ^{\prime} )}
g_{\phantom{\mu} \nu }^{\mu}\left( z,z^{\prime }\right) ,  \label{RedGreenDef}
\end{equation}%
where $ \textbf{R}= (x-x^{\prime },\;y-
y^{\prime })$ and  $\mathbf{p}=(p_{x},p_{y})$ is the
momentum parallel to $\Sigma$
\cite{Schwinger}. In Eq.
(\ref{RedGreenDef}) we have omitted 
the dependence of the reduced GF
$g_{\phantom{\mu} \nu }^{\mu }$ on $\omega$ and $
\mathbf{p}$. Due to the antisymmetry of the
Levi-Civit\`a symbol, the partial derivative
appearing in the second term of 
Eq. (\ref{EqsGreenMatrix}) does not
introduce  derivatives with respect to $z
$. This allows us to write the equation for the
reduced GF $g_{\phantom{\nu} \sigma }^{\nu }
( z,z^{\prime })$ as
\begin{equation}
\left[ \eta _{\phantom{\mu}\nu }^{\mu } \partial ^{2} + i
\tilde{\theta}\delta \left( \Sigma \right)
\epsilon _{\phantom{3\mu \alpha} \nu }^{3 \mu \alpha}
p_{\alpha }\right] g_{\phantom{\nu} \sigma }^{\nu }
\left( z,z^{\prime }\right) =
\eta _{\phantom{\mu} \sigma }^{\mu}
\delta \left( z-z^{\prime }\right) ,  \label{RedGreenFunc}
\end{equation}
where now $\partial ^{2}= \mathbf{p}
^{2} - \omega ^{2} - \partial _{z}^{2}$
and $p^{\alpha}=\left( \omega ,
\mathbf{p} , 0 \right) $.  In solving Eq. (\ref{RedGreenFunc})  we
employ a method similar to that used for
obtaining the GF for the one-dimensional
$\delta$-function potential in quantum
mechanics, where the free GF is used for
integrating the GF equation with $\delta
$-interaction. 
Here 
the free GF  we  use
is the reduced GF for two parallel
conducting surfaces placed at
$z=0$ and $z=L$, which is the solution of
$\partial ^{2} \mathfrak{g} \left( z ,
z^{\prime} \right) = \delta \left( z - z
^{\prime} \right)$ satisfying the BC
$\mathfrak{g} \left(
0 , z^{\prime} \right) = \mathfrak{g}
\left( L , z^{\prime} \right) = 0$, namely:
\begin{equation}
\mathfrak{g} \left( z , z ^{\prime} \right) = \frac{\sin
\left[ p z _{<} \right] \sin \left[ p \left( L - z _{>} \right) \right]}%
{p \sin \left[ p L \right]} , \label{RGF-plates}
\end{equation}
where $z _{>}$ ($z _{<}$) is the greater
(lesser) of $z$ and $z ^{\prime}$, and $p
= \sqrt{\omega ^{2} - \textbf{p} ^{2}}$.
Now, Eq. (\ref{RedGreenFunc}) can
be directly integrated using the
free GF (\ref{RGF-plates}) together
with
the properties of the $\delta$-function,
reducing the problem to a set of
coupled algebraic equations,
\begin{equation}
g ^{\mu} _{\phantom{\mu} \sigma} \left( z , z ^{\prime} \right) =
\eta ^{\mu} _{\phantom{\mu} \sigma} \mathfrak{g}
\left( z , z ^{\prime} \right) - i \tilde{\theta}
\epsilon ^{3 \mu \alpha} _{\phantom{3 \mu \alpha} \nu }
p_{\alpha } \mathfrak{g} \left( z , a \right)
g ^{\nu} _{\phantom{\nu} \sigma} \left( a , z ^{\prime} \right) .  \label{A3}
\end{equation}
Note that the continuity of
$\mathfrak{g}$ at $z=z^{\prime}$
implies the continuity of $g ^{\mu} _{\phantom{\mu} \sigma}$  there,
and the discontinuity
of $\partial _{z} \mathfrak{g}$ at the same
point yields the corresponding discontinuity of
$\partial_{z} g ^{\mu} _{\phantom{\mu} \sigma}$, in
accordance with the BC (\ref{BC4Pot}).
We write the general solution to Eq. (\ref{A3})  as the sum of two terms,
$g^{\mu} _{\phantom{\mu}\nu } \left( z , z ^{\prime} \right) =
\eta ^{\mu } _{\phantom{\mu} \nu } \mathfrak{g}
\left( z , z ^{\prime } \right) + g ^{\mu} _{\theta \nu}
\left( z , z^{\prime} \right) $.
The first term provides the propagation in
the absence of the TI.
The second, 
to be called  the reduced $\theta$-GF, which 
can be shown to be
\begin{eqnarray}
g ^{\mu} _{\theta \nu} \left( z , z^{\prime} \right)=
\tilde{\theta} \mathfrak{g} \left( a , a \right) \left[ p^{\mu} p_{\nu}-
\left( \eta ^{\mu} _{\phantom{\mu} \nu } + n ^{\mu} n _{\nu} \right)
p ^{2} \right] P \left( z , z^{\prime} \right) \nonumber \\ +\, i \,
\epsilon ^{\mu \phantom{\nu} \alpha 3} _{\phantom{\mu} \nu }
p_{\alpha } P \left( z , z^{\prime} \right) , \label{theta-GF}
\end{eqnarray}
encodes the TME due to the
topological $\theta$-term. Here
\begin{equation}
P \left( z,z^{\prime }\right) = - \tilde{\theta} \frac{\mathfrak{g}
\left( z , a \right) \mathfrak{g}\left( a,z^{\prime }\right) }%
{1- p^{2} \tilde{\theta}^{2}\mathfrak{g}^{2}\left( a,a\right) }.
\label{A(Z,Z)}
\end{equation}
In the static limit ($\omega = 0$), our result (\ref{theta-GF})
reduces to the one reported in Ref.
\cite{MCU}. Clearly,  the full GF matrix  $G ^{\mu} _{\phantom{\mu} \nu} \left( x , x ^{\prime} \right)$  can also be
written as the sum of two terms,
$G ^{\mu} _{\phantom{\mu} \nu} \left( x , x ^{\prime} \right) =
\eta ^{\mu} _{\phantom{\mu}\nu}
\mathcal{G} \left( x , x ^{\prime} \right) +
G ^{\mu} _{\theta \nu} \left( x , x ^{\prime} \right) $. We
remark in passing that the reciprocity
relation for the GF, $G_{\mu \nu} ( x , x
^{\prime }) = G_{\nu \mu }(x
^{\prime},x)$, is a direct consequence of the property $g _{\mu
\nu} \left( z , z ^{\prime} , p ^{\alpha}
\right) = g _{\nu \mu}\left( z ^{\prime} ,
z , - p ^{\alpha} \right)$.
\section{III. The vacuum stress-energy tensor}
\label{vacuum}
In the previous section we showed that
the $\theta$-term modifies the behaviour
of the fields at the surface $\Sigma $
only. This suggests that, for the bulk
regions, the stress-energy tensor
(SET) for $\theta $-ED has the same form
as that in standard electrodynamics. 
In fact, in Ref. \cite{MCU} we explicitly computed  
the SET and verified the latter. As it turns out
the SET can be cast in the from:
\begin{equation}
T ^{\mu \nu} = \frac{1}{4 \pi} \left( - F ^{\mu \lambda}
F ^{\nu} _{\phantom{\nu}\lambda} + \frac{1}{4}
\eta^{\mu \nu} F _{\alpha \beta} F ^{\alpha \beta} \right) .
\label{Stress-Energy}
\end{equation}
Clearly this tensor is traceless, \textit{i.e.,}
$T ^{\mu} _{\phantom{\mu} \mu} = 0$ and its divergence
is
\begin{equation}
\partial _{\mu} T ^{\mu \nu} = - F ^{\nu} _{\phantom{\nu} \lambda}
j ^{\lambda} - ( \tilde{\theta} / 4 \pi ) \delta \left( \Sigma \right)
n _{\mu} F ^{\nu} _{\phantom{\nu} \lambda}
\tilde{F} ^{\mu \lambda} . \label{DivST}
\end{equation}
As expected, the SET
it is not conserved at $\Sigma$
because of the TME which induces effective charge
and current densities there.

Now we address the
vacuum expectation
value of the  SET,  to
which we will refer simply as the vacuum
stress (VS).
The local approach to compute the VS
was initiated by Brown and Maclay who
calculated the renormalised stress tensor  by
means of GF techniques
\cite{Brown,Schwinger1}. In there, the
VS can be obtained from appropriate derivatives of the GF, in virtue of Eq. (2.1E) from Ref. 
\cite{Candelas},
\begin{equation}
G ^{\mu \nu} \left( x , x ^{\prime} \right) = - i \left< 0 \right|
\hat{\mathcal{T}}  A ^{\mu} \left( x \right) A ^{\nu}
\left( x ^{\prime} \right)  \left| 0 \right> . \label{VacuumGreen}
\end{equation}
Using the standard point splitting technique and  taking the vacuum expectation value of the SET
(\ref{Stress-Energy}) we find
\begin{eqnarray}
\left<  T ^{\mu \nu} \right> = \frac{i}{4 \pi} \lim _{x \rightarrow x ^{\prime}}
\Big[ - \partial ^{\mu}  \partial ^{\prime \nu}
G ^{\lambda} _{\phantom{\lambda} \lambda} + \partial ^{\mu}
\partial _{\lambda} ^{\prime} G ^{\lambda \nu} +
\partial ^{\lambda} \partial ^{\prime \nu} G ^{\mu} _{\phantom{\mu} \lambda}
\nonumber \\
-\,  \partial ^{\prime \lambda} \partial _{\lambda} G ^{\mu \nu} + \frac{1}{2}
\eta ^{\mu \nu} \left( \partial ^{\alpha} \partial _{\alpha} ^{\prime}
G ^{\lambda} _{\phantom{\lambda} \lambda} -
\partial ^{\alpha} \partial _{\beta} ^{\prime}
G ^{\beta} _{\phantom{\beta} \alpha} \right) \Big] ,
\label{VacuumStress}
\end{eqnarray}
where we have omitted the
dependence of $G^{\mu \nu}$ on
$x$ and $x ^{\prime}$. This result can be further simplified
as follows. Since the GF is written as the sum of two terms,
then the VS can
also be written in the same way, \textit{i.e.,}
\begin{equation}
\left< T ^{\mu \nu} \right> = \left< t ^{\mu \nu} \right> + %
\left< T ^{\mu \nu} _{\theta} \right>.
\end{equation}
The first term,
\begin{equation}
\left< t ^{\mu \nu} \right> = \frac{1}{4 \pi i} \lim _{x \rightarrow x ^{\prime}}
\left( 2 \partial ^{\mu} \partial ^{\prime \nu} - \frac{1}{2} \eta ^{\mu \nu}
\partial ^{\lambda} \partial _{\lambda} ^{\prime} \right) \mathcal{G}
\left( x , x ^{\prime} \right) , \label{SE-free}
\end{equation}
is the VS in the absence of the
TI. In obtaining Eq.
(\ref{SE-free}) we use that the
GF is diagonal when the TI is absent, \textit{i.e.} it is equal to
$\eta^{\mu} _{\phantom{\mu} \nu} \mathcal{G}
\left( x , x^{\prime} \right)$. The second term, to
which we will refer as the
$\theta$ vacuum stress ($\theta$-VS), can be
simplified since the $\theta$-GF satisfies
the Lorentz gauge condition $\partial _{\mu} G
^{\mu \nu} _{\theta} = 0 $. The
proof follows from the
reduced GF of Eq. (\ref{theta-GF}):
\begin{equation}
\partial _{\mu} G ^{\mu} _{\theta \nu} \left( x , x ^{\prime} \right)
\propto \int \int p _{\mu} g _{\theta \nu }^{\mu}
\left(z , z^{\prime} \right) , \label{LorentzGreen}
\end{equation}
which vanishes given that
$\epsilon _{\phantom{\mu} \nu }^{\mu \phantom{\nu} \alpha 3}
p _{\mu} p_{\alpha } = 0$ and $p _{\mu} n ^{\mu} = 0$.
With the previous result
the $\theta$-VS can be written as
\begin{equation}
\left< T ^{\mu \nu} _{\theta} \right> = \frac{1}{4 \pi i}
\lim _{x \rightarrow x ^{\prime}}  \left[ \partial ^{\mu}
\partial ^{\prime \nu} G _{\theta} + \partial ^{\prime \lambda}
\partial _{\lambda} \left( G ^{\mu \nu} _{\theta} - \frac{1}{2}
\eta ^{\mu \nu} G _{\theta} \right) \right] , \label{VacuumStress2}
\end{equation}
where $G _{\theta} = G ^{\mu} _{\theta
\mu}$ is the trace of the $\theta$-GF.
This result exhibits the
vanishing of the trace at quantum level,
\textit{i.e.,} $\eta _{\mu
\nu} \left<  T ^{\mu \nu} _{\theta} \right> = 0
$. For the simplest situation in which the
SET is conserved, one can verify that $\partial _{\nu}
\left< T ^{\mu \nu} \right> = \left<
\partial _{\nu} T ^{\mu \nu} \right>=0$.
However, as pointed out by Deutsch and
Candelas \cite{Candelas}, this identity
need not be a rule.
The problem at hand is an example of this,
since $ \left< \partial _{\nu} T ^{\mu \nu} _{\theta} \right> = ( \tilde{\theta} / 8
\pi i ) \delta (\Sigma) \eta ^{\mu 3} \epsilon ^{\sigma \nu  \alpha \beta}
\lim _{x \rightarrow x ^{\prime}} \partial_{\sigma} \partial ^{\prime} _{\alpha} G
_{\theta \nu \beta}$,
which can be shown to be
different from $\partial _{\nu} \left< T
_{\theta} ^{\mu \nu} \right>$.

\section{IV. The Casimir Effect}

\label{casimir-sec}

 Now we consider  the problem of calculating the
renormalised VS $\left< T^{\mu \nu} \right>
_{\mathrm{ren}}$, which
is obtained
as the difference
between the VS in the
presence of boundaries and that of the
free vacuum.
In the standard case
($\theta =0$),
Brown and Maclay \cite{Brown} obtained
that it is uniform between the plates,
\begin{equation}
\left< t^{\mu \nu} \right>_{\mathrm{ren}} = - \frac{\pi ^{2}}{720 L ^{4}}
\left( \eta ^{\mu \nu} + 4 n ^{\mu} n ^{\nu} \right) , \label{BM-T}
\end{equation}
with $L$ the distance between the plates.
The Casimir stress
on the plates  was obtained by
differentiating the Casimir energy
${\mathcal{E}}_L = L \left< t^{00} \right> _{\mathrm{ren}} =
- \pi ^{2} /720 L ^{3}$ with respect to $L$, \textit{i.e.,}
$F_L = - d {\mathcal{E}}_L / dL = - \pi ^{2} /240 L ^{4}$.

Now our concern is to calculate the 
renormalised
$\theta$-VS (\ref{VacuumStress2})
for our Casimir system. We proceed along
the lines of Refs. \cite{Brown, Candelas}.
From Eq. (\ref{VacuumStress2}),
together with the symmetry of the
problem we find that  the 
 $\theta$-VS  can be written as
\begin{eqnarray}
\left< T ^{\mu \nu} _{\theta} \right> = i \tilde{\theta} \int
\frac{d ^{2} \textbf{p}}{(2 \pi) ^{2}} \int \frac{d \omega}{2 \pi}
\left( p ^{\mu} p ^{\nu} + n ^{\mu} n ^{\nu} p ^{2} \right)  \times \nonumber \\
\mathfrak{g} \left( a , a \right)
\lim _{z \rightarrow z ^{\prime}}  \left( p ^{2} +
\partial ^{\prime} _{z} \partial _{z} \right)
P \left( z , z ^{\prime} \right) ,  \label{VS2}
\end{eqnarray}
where we have used $\partial ^{\mu} = (- i \omega
, i \textbf{p} , \partial _{z})$. In deriving
this result we used the Fourier
representation of the GF in Eq. 
(\ref{RedGreenDef}) together with the
solution for the reduced $\theta$-GF
given by Eq. (\ref{theta-GF}). From the result (\ref{VS2}) we calculate the
renormalised  $\theta$-VS, which is
given by $\left< T^{\mu \nu}
_{\theta} \right> _{\mathrm{ren}} = \left< T ^{\mu \nu}
_{\theta} \right> - \left< T
^{\mu \nu} _{\theta} \right>
_{\mathrm{vac}}$, where the  first (second)
term is the $\theta$-VS in the
presence (absence) of the plates  \cite{Candelas}.
When the plates are absent, the reduced
GF we have to use to compute the $
\theta$-VS in the region $[0,L]$ is that
of the free-vacuum, $ \mathfrak{g} _{0}
\left( z , z ^{\prime} \right) = (i / 2p )
\exp ( ip \vert z - z ^{\prime} \vert ) $
\cite{Schwinger}, from
which we find that $ \lim
_{z \rightarrow z ^{\prime}} \partial _{z}
\partial _{z} ^{\prime} P _{0} \left( z , z
^{\prime} \right) = - p ^{2} \lim _{z
\rightarrow z ^{\prime}} P _{0} \left( z ,
z ^{\prime} \right) $, thus implying that
the integrand in Eq. (\ref{VS2}) vanishes.  The function
$P _{0}$ is given by  Eq.
(\ref{A(Z,Z)}) when the free-vacuum
reduced GF $\mathfrak{g}(z,z^{\prime})$ is replaced by  $\mathfrak{g}_0(z,z^{\prime})$.
Therefore we conclude that $ \left< T
^{\mu \nu} _{\theta} \right> _{vac} =
0$.

\begin{figure}[tbp]
\begin{center}
\includegraphics[scale=0.69]{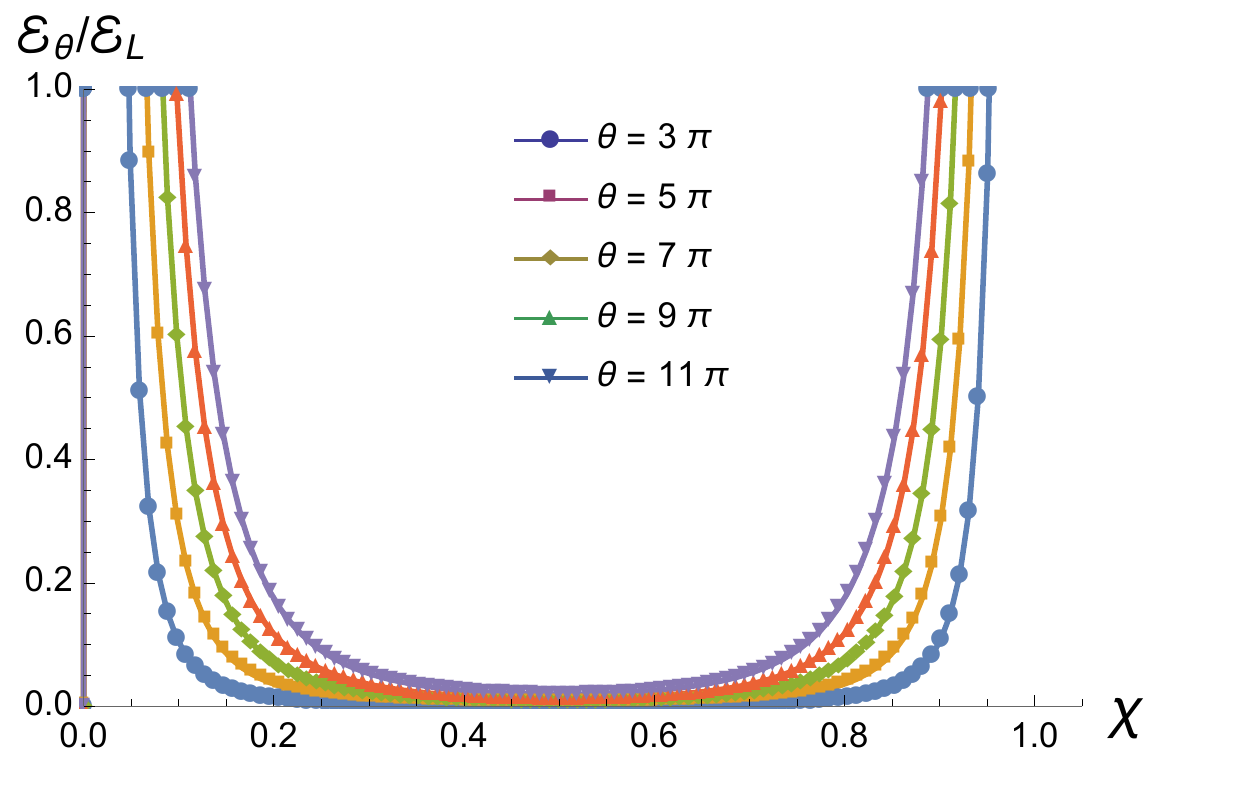}
\end{center}
\caption{{\protect\small The ratio $
\mathcal{E}_{\theta} / \mathcal{E}_{L}$ as
a function of the dimensionless distance $
\chi = a / L$, for different values of $
\theta$. }}
\label{FigCasEnergy}
\end{figure}

Next we compute $\left< T ^{\mu \nu}_{\theta} \right> _{\mathrm{ren}}=\left< T ^{\mu \nu} _{\theta} \right>$ starting from Eq. (\ref{VS2}). From the
symmetry of the problem,
the components of the stress
along the plates, $\left< T ^{11}
_{\theta} \right>$ and $
\left< T ^{22} _{\theta} \right>
 $, are equal. In addition, from the mathematical
structure of Eq. (\ref{VS2}) we find the
relation $\left< T ^{00}
_{\theta} \right> = - \left< T ^{11} _{\theta}
\right> $.
These results, together with the traceless
nature of the SET, allow us to write the
renormalised $\theta$-VS in
the form
\begin{equation}
\left< T ^{\mu \nu} _{\theta} \right> _{\mathrm{ren}} = \left( \eta ^{\mu \nu} + 4 n ^{\mu} n ^{\nu} \right) \tau ( \theta , z)  \; , \label{VS3}
\end{equation}
where
\begin{eqnarray}
\tau( \theta , z) &=& i \tilde{\theta} \int \frac{d ^{2} \textbf{p}}{(2 \pi) ^{2}} \int \frac{d \omega}{2 \pi} \omega ^{2} \mathfrak{g} \left( a , a \right) \times \nonumber \\
&& \lim _{z \rightarrow z ^{\prime}}  \left( p ^{2} + \partial ^{\prime} _{z} \partial _{z} \right) P \left( z , z ^{\prime} \right) . \label{f-func}
\end{eqnarray}
Note that our $\theta$-VS exhibits the
same tensor structure as  the result
obtained by Brown and Maclay (\ref{BM-T}), 
but we obtain a $z$-dependent VS since
the SET is not conserved at $\Sigma$.
To understand better 
$\tau (\theta , z)$, we compute the
limit of the integrand {in Eq. (\ref{f-func})}. Using Eq.
(\ref{RGF-plates})
we obtain
\begin{eqnarray}
\lim _{z \rightarrow z ^{\prime}} \left( p ^{2} + \partial _{z} \partial _{z} ^{\prime} \right) P \left( z , z ^{\prime} \right) = - \frac{\tilde{\theta}}{1 - \tilde{\theta} ^{2} p ^{2} \mathfrak{g} ^{2} \left( a , a \right)} \times \nonumber \\ \left\lbrace \frac{\sin ^{2} \left[p \left( L - a \right) \right]}{\sin ^{2} \left[p L \right]} H \left( a - z \right) + \frac{\sin ^{2} \left[ p a \right]}{\sin ^{2} \left[p L \right]} H \left( z - a \right) \right\rbrace. \label{Derivatives}
\end{eqnarray}
{To evaluate the integral in Eq.~(24) 
we first write the momentum element as $d ^{2} \textbf{p} = \vert \textbf{p} \vert d \vert \textbf{p} \vert d \vartheta$ and
integrate $\vartheta$. Next, we
perform a Wick rotation such that $\omega \rightarrow i \zeta $, then replace $\zeta$ and $\vert \textbf{p} \vert$ by plane polar coordinates $\zeta = \xi \cos \varphi$, $\vert \textbf{p} \vert= \xi \sin \varphi$ and finally integrate
$\varphi$. } The renormalised $\theta$-VS in Eq.~(\ref{VS3})
then becomes
\begin{eqnarray}
\left< T ^{\mu \nu}_{\theta} \right>_{\mathrm{ren}}=- \frac{\pi ^{2}}{720 L ^{4}} \left( \eta ^{\mu \nu} + 4 n ^{\mu} n ^{\nu} \right)  \times \nonumber \\  \left[ u ( \theta , \chi ) H \left( a - z \right) +  u ( \theta , 1 - \chi ) H \left( z - a \right) \right] , \label{VS4}
\end{eqnarray}
where
\begin{equation}
u( \theta , \chi ) = \frac{120}{\pi ^{4}} \int _{0} ^{\infty}  \frac{\tilde{\theta} ^{2} \xi ^{3} \mbox{\small sh} \left[ \xi \chi \right] \mbox{\small sh}^{3} \left[ \xi \left( 1 - \chi \right) \right] \mbox{\small sh}^{-3} \left[ \xi \right]}{1 + \tilde{\theta}^{2} \mbox{\small sh} ^{2} \left[ \xi \chi \right] \mbox{\small sh}^{2} \left[ \xi \left( 1 - \chi \right) \right] \mbox{\small sh}^{-2} \left[ \xi \right] } d \xi , \label{funcionU}
\end{equation}
with $\mbox{\small sh}(x) = \sinh (x)$ 
and $\chi = a / L$ with $0 < \chi
< 1$. 
Physically, we interpret the function
$u( \theta , \chi)$ as the ratio
between the renormalised
$\theta$-energy density in the vacuum
region $[0 , a )$ and that of the
renormalised energy density in the
absence of the TI, $
\left< t ^{00} \right> _{\mathrm{ren}}$.
The function $u ( \theta , 1 - \chi
)$ has an analogous interpretation for the
bulk region of the TI $(a , L]$. 
This shows that the energy
density is constant in the bulk regions,
however a simple
discontinuity arises at $\Sigma$, \textit{i.e.,}
$\partial _{z} \left< T ^{00}
_{\theta} \right> _{\mathrm{ren}} \propto \delta (\Sigma)$.
The Casimir energy  ${\cal E}={\cal E}_L+{\cal E}_{\theta}$ is defined as the
energy per unit of area stored in the
electromagnetic field between the plates.
To obtain it we must integrate the contribution from 
the
$\theta$-energy  density. 
\begin{equation}
\mathcal{E}_{\theta}=\int _{0} ^{L} dz \left< T ^{00} _{\theta} \right> _{\mathrm{ren}} = {\mathcal{E}}_L \left[ \chi  u ( \theta , \chi ) + (1 - \chi)  u ( \theta , 1 - \chi ) \right], \label{CasEnergy}
\end{equation}
recalling that $\mathcal{E}_L$ is the Casimir energy
in the absence of the TI. The first term
corresponds to the energy
stored in the electromagnetic field
between $P _{1}$
and $ \Sigma$, while the second term is
the energy stored in the bulk of the TI.
The ratio $
\mathcal{E} _{\theta} / {\mathcal{E}}_L$ as
a function of $\chi$ for different values of $
\theta$ (appropriate for TIs {\cite{Grushin}}) is 
plotted 
in Fig. \ref{FigCasEnergy}.

\subsection{IV-a The Casimir stress on the $\theta$-piston}

\begin{figure}[tbp]
\begin{center}
\includegraphics[scale=0.69]{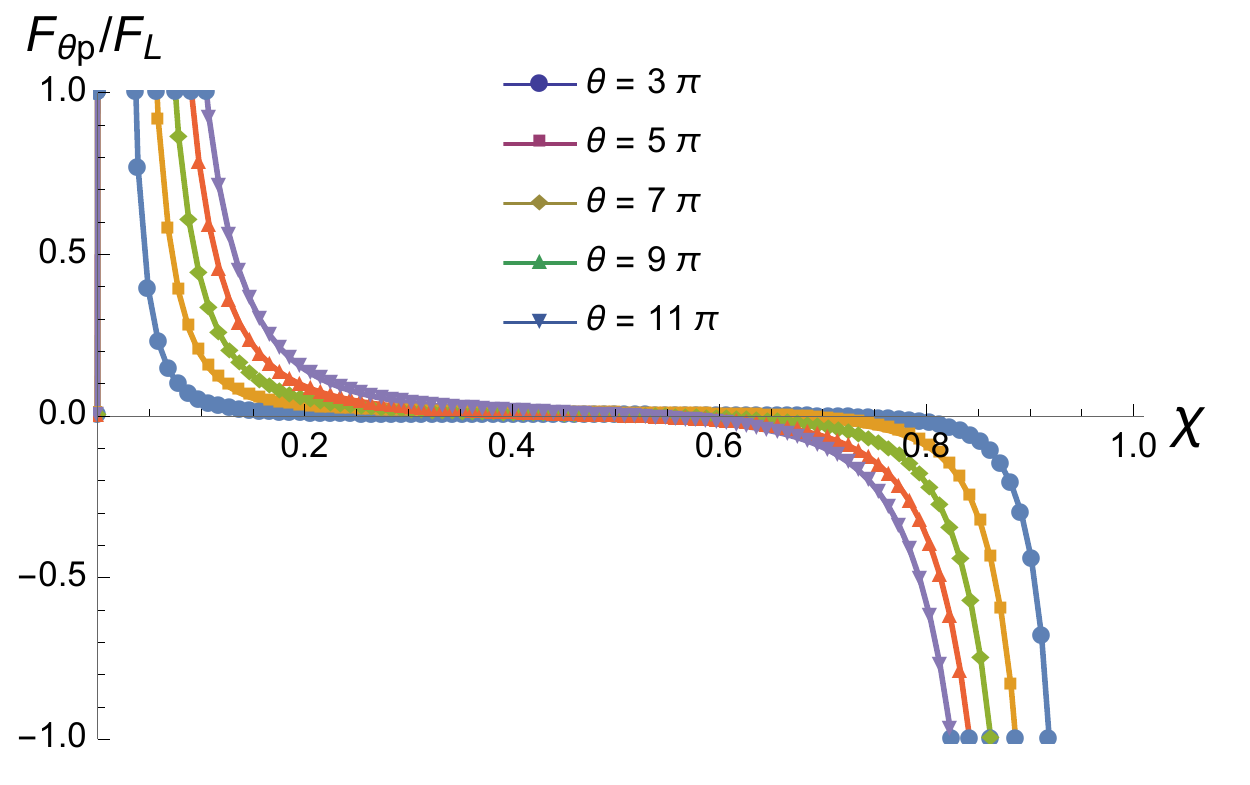}
\end{center}
\caption{{\protect\small  The
Casimir stress on the $\theta$-piston  in units of $F_L$
as a function of $\chi$, for different
values of $\theta$}.}
\label{ForcePiston}
\end{figure}
The setup known in the literature as the Casimir piston consists of a
rectangular box of length $L$ divided by a
movable mirror (piston) at a distance $a$
from one of the plates \cite{Cavalcanti}.
 The net result is
that the Casimir energy  in each region generates a force
on the piston pulling it towards the
nearest end of the box. Here we consider
a similar setup, which we have called the
$\theta$-piston, in which the piston is 
the 
TI.
Since the surface $\Sigma$ changes
the energy density of the electromagnetic
field in the bulk regions,   an
effective Casimir stress acts upon
$\Sigma$. This can be obtained as 
$F_{\theta p} = - d
\mathcal{E} _{\theta} / da$. The result is
\begin{equation}
\frac{F _{\theta p} }{F_L} = - \frac{1}{3} \frac{d}{d \chi} \left[ \chi  u ( \theta , \chi ) + (1 - \chi)  u ( \theta , 1 - \chi ) \right] ,
\end{equation}
where $F_L $ is
the Casimir stress between the two
perfectly reflecting plates in the absence
of the TI.
Figure \ref{ForcePiston} shows the
Casimir stress on $\Sigma$ in units of $F_L$
as a function of $\chi$ for different
values of $\theta$. We observe that this force
pulls the boundary $\Sigma$ towards the closer of the two
fixed walls $P_1$ or $P_2$, similarly to the conclusion in Ref.\cite{Cavalcanti}.
\subsection{IV-b Casimir stress between $P _{1}$ and $\Sigma$, when $L \rightarrow \infty$}
Now let us consider the limit where
the plate $P_{2}$ is sent 
to infinity, 
\textit{i.e.,} $L
\rightarrow \infty$. This
configuration corresponds to a perfectly
conducting plate $P_1$ in vacuum, and
a semi-infinite TI located at a
distance $a$. Here the plate and the
TI exert a force upon each
other. The Casimir energy
(\ref{CasEnergy}) in the limit $L
\rightarrow \infty$ takes the form 
$\mathcal{E}_{\theta} ^{L \rightarrow
\infty} = {\mathcal{E}}_a R ( \theta)$, 
with ${\mathcal{E}}_a= - \pi^{2} / 720 a ^{3}$, and the function
\begin{equation}
R (\theta) = \frac{120}{\pi ^{4}} \int _{0} ^{\infty} \xi ^{3} \frac{\tilde{\theta} ^{2}}{1 + \tilde{\theta} ^{2} e ^{- 2 \xi } \sinh ^{2} \xi} e ^{- 3 \xi } \sinh \xi d \xi ,
\end{equation}
is $a$-independent  and  bounded
by its  $\theta \rightarrow
\pm \infty$ limit, \textit{i.e.,}
\begin{equation}
R (\theta ) \leq \frac{120}{\pi ^{4}} \int _{0} ^{\infty} \xi ^{3} \frac{e ^{ - \xi }}{\sinh \xi}  d \xi = 1 .
\end{equation}
Thus, for this case,  the energy stored
in the electromagnetic field
is bounded
by the Casimir energy
between two parallel conducting plates at a distance $a$,
\textit{i.e.,} 
$\mathcal{E} _{\theta} ^{L \rightarrow
\infty} \leq {\mathcal{E}}_a$. Physically this
implies that in the  $\theta \rightarrow
\infty $ limit the surface of the TI
mimics a conducting plate, which is analogous to
Schwinger's prescription for describing a
conducting plate as the  $
\varepsilon \rightarrow \infty$ limit of 
material media \cite{Schwinger}. 
{These  results, which stem from our
Eqs.~(\ref{theta-GF}) and (\ref{A(Z,Z)}), 
agree 
with those obtained in the 
global 
energy approach which uses the reflection matrices 
containing the Fresnel
coefficients as in Ref. \cite{Grushin}, 
when 
the 
appropriate limits 
to describe an ideal conductor at $P_1$ and a purely topological surface at
 $\Sigma$ are taken 
into account. The calculation, however, is  too long to be shown  here.} 
Taking
the derivative with respect to $a$ we find
that the plate and the TI
exert a force (in units of $F_a = - \pi ^{2}
/ 240 a ^{4}$) of attraction upon each
other
given by $f _{\theta} = F ^{L \rightarrow
\infty} _{\theta} / F_a = R ( \theta )$. 
Numerical results for  $f_{\theta}$ for different values of $\theta$ are presented in Table \ref{table1}.
\begin{table}
\centering
\begin{tabular}{ c | c  c  c  c c }
$\theta$ & $\pm 7 \pi$ & $\pm 15 \pi$ & $\pm 23 \pi$ & $\pm 31 \pi$ & $\pm 39 \pi$ \\
\hline
$f _{\theta}$ & 0.0005 & 0.0025 & 0.0060 & 0.0109 & 0.0172 \\
\end{tabular}.
\caption{{\protect\small Normalized force $f_{\theta}=F ^{L \rightarrow
\infty} _{\theta} / F_a = R ( \theta )$ for different values of $\theta$.}}
\label{table1}
\end{table}
\section{V. Conclusions and Outlook}
In this letter we have used 
the stress-energy tensor to compute  the 
Casimir energy and stress for the system  shown in 
Fig. \ref{fig1}. The setup consists of a slab of a coated 
topological insulator (TI) of constant  topological 
magnetoelectric  polarisability $\theta$ in the 
region $a<z<L$, which  partially fills 
the space between two perfectly reflecting parallel plates. 
In a first approach, we
ignore features that are relevant in
experimental situations, such as the
optical properties of TIs and  temperature
effects,  
{which can be taken into account including these parameters in the Green's function according
to standard methods Refs.~\cite{Milton,Bordag,Schwinger}}. 
The system is well described by the action of 
$\theta$ electrodynamics given in Eq.~(\ref{action}). 
In this work, we obtained the renormalised vacuum 
stress from (derivatives of) the corresponding Green's  
function (GF) of the problem.  To this end we require the 
time dependent GF, for which we  extended our results of  
Ref.~\cite{MCU}, which dealt with the static case. This GF can be exactly calculated because the TI introduces a 
localised discontinuity in the equations of motion along the 
$z$ direction, while the dependence in time and in the remaining coordinates is invariant under the respective translations. We  considered two cases: (1) the $\theta$-piston, defined in the interval $0<z<L$. The  contribution to the Casimir energy ${\mathcal E}={\mathcal E}_L+{\mathcal E}_\theta$, that arises from the TI, ${\mathcal E}_\theta$,  is plotted in Fig.~\ref{FigCasEnergy}, in units of ${\mathcal E}_L$, as a function of the reduced length $\chi$ and for various values of $\theta$. The  Casimir stress on the boundary $\Sigma$, $F_{\theta p}=-d\mathcal{E}/d a$, is plotted in Fig. \ref{ForcePiston}, in units of $F_L$. We observe that this force
pulls  $\Sigma$ towards the closer of the two
fixed walls $P_1$ or $P_2$, similarly to the conclusion in Ref.\cite{Cavalcanti}.
(2) The second case we have considered is the  
$L \rightarrow \infty$ limit of the $\theta$-piston, which describes the interaction, via the interface $\Sigma$, between the conducting plate $P_1$ in vacuum and  a semi-infinite TI located at a distance $a$. The corresponding Casimir stress, in units of $F_a$, is shown in Table \ref{table1}, for
different values of $\theta$.
{These results, which rely
on the due evaluation of GFs and the renormalised 
stress-energy tensor, are in perfect 
agreement with those obtained following the 
global energy approach of Ref. \cite{Grushin}.
We also remark
that the  discontinuity of the vacuum expectation 
value of the energy density obtained
in Eq. (\ref{VS4}) is finite, similar to that
in Ref. \cite{Milton2}.
Although,  in our case, a physical interpretation
of such discontinuity  is not
immediate, it is somehow expected
due to the non-conservation of the stress-energy 
tensor at the $\Sigma$ boundary owing to the self-induced
charge and current densities there \cite{MCU}.
}

A general feature of our analysis is that the  TI
induces a $\theta$-dependence on the Casimir stress, which 
could be used to measure $\theta$. 
Since  the Casimir stress  has been measured for 
separation distances 
in the $0.5- 3.0\,\mu$m range \cite{Bressi}, these 
measurements  require
TIs of width lesser than $0.5 \mu$m and an increase of the  experimental precision of two to three orders of magnitude.
In  practice the measurability of $f_\theta$ 
depends on the value of the
TMEP, which is
quantised as
$\theta = (2n + 1) \pi, \, n \in
\mathbb{Z}$.
The particular values $\theta = \pm 7 \pi
, \pm 15 \pi$ are appropriate for the TIs
such as Bi$_{1-x}$Se$_{x}$ 
{\cite{Zhou}}, where
we have  $f _{\pm 7 \pi}
\approx 0.0005$ and $f _{\pm 15 \pi}
\approx 0.0025$, which are not yet
feasible with the present experimental
precision. 
This
effect could also be explored in  TIs
described by a higher  coupling $\theta$, 
such as Cr$_{2}$O$_{3}$. 
However, this
material induces more
general magnetoelectric couplings   not
considered in 
our model \cite{Grushin}.

{Although the reported $\theta$-effects of our Casimir systems cannot be  observed 
in the  laboratory yet, 
we have aimed
to establish the Green's function method as an alternative  theoretical framework
for dealing with the topological magnetoelectric effect of TIs
and also as yet another application of the GF method we
developed in Ref. \cite{MCU}. 

Though not  frequently used in the corresponding TIs literature, this method  plays an important role in the calculations of the standard Casimir effect, besides its usefulness in many other topics in $\theta$ electrodynamics. }

\acknowledgments
 LFU has been supported in part by the projects DGAPA(UNAM) IN104815 and CONACyT (M\'{e}xico) 237503. M. Cambiaso has been supported in part by the project FONDECYT (Chile) 11121633.

\end{document}